\documentclass[superscriptaddress,aps,prl,twocolumn,amsmath,amssymb]{revtex4-1}

\usepackage{graphicx}
\usepackage{dcolumn}
\usepackage{color,soul}
\usepackage[colorlinks=true, citecolor=red, urlcolor=blue]{hyperref}

\usepackage{color}

\begin{document}

\title{Metastable rocksalt ZnO is $p$-type dopable}

\author{Anuj Goyal}
\affiliation{Colorado School of Mines, Golden, CO 80401, USA}
\affiliation{National Renewable Energy Laboratory, Golden CO 80401, USA}

\author{Vladan Stevanovi\'{c}}
\email{vstevano@mines.edu}
\affiliation{Colorado School of Mines, Golden, CO 80401, USA}
\affiliation{National Renewable Energy Laboratory, Golden CO 80401, USA}



\begin{abstract}
%
Despite decades of efforts, achieving $p$-type conductivity in the wide band gap ZnO in its ground-state wurtzite structure continues to be a challenge. Here we detail how $p$-type ZnO can be realized in the metastable, high-pressure rocksalt phase (also wide-gap) with Li as an external dopant. 
Using modern first-principles defect theory, we predict Li to dope the rocksalt phase $p$-type by preferentially substituting for Zn and introducing shallow acceptor levels. Formation of compensating donors like interstitial Li and/or hydrogen, ubiqutous in the wurtzite phase, is inhibited by the close-packed nature of the rocksalt structure, which also exhibits relatively high absolute valence band edge that promotes low hole effective mass and hole delocalization. Resulting concentrations of free holes are predicted to exceed $\sim10^{19}$ cm$^{-3}$ under O-rich synthesis conditions while under O-poor conditions the system remains $n$-type dopable. In addition to revealing compelling opportunities offered by the metastable rocksalt structure in realizing a long-sought $p$-type ZnO our results present polymorphism as a promising route to overcoming strong doping asymmetry of wide-band gap oxides.

\end{abstract}


\maketitle

\section{\label{sec:I}Introduction}
Absence of a high-mobility $p$-type transparent semiconductor material represents a critical bottleneck to advancing present optoelectronic devices and realizing active bipolar transparent electronics \cite{Thomas1997, Kawazoe1997, Wager2003, Wager2008, Hosono2011a}. Strong proclivity of wide band gap semiconductors, oxides in particular, to $n$-type conductivity, limits current transparent oxide applications to unipolar and/or passive \cite{Wager2008, Hosono2011a}. Following the success of GaN \cite{Nakamura2015}, significant efforts have been made in realizing $p$-type ZnO; however, only with limited success \cite{Ozgur2005}. Despite its practical advantages over GaN, due to natural abundance and the more facile synthesis, ZnO in its ground state wurtzite structure strongly opposes $p$-type doping. This behavior of ZnO is well understood and follows mainly from the relatively low position of its band edges on the absolute energy scale that promotes both localization of holes (formation of deep acceptor states) and the pronounced donor behavior of ubiquitous impurities such as interstitial hydrogen \cite{Anderson2009, VanDeWalle2000a}. 
The external $p$-type dopants, such as the group-V (N, P, As) substitutes for oxygen or group-I (Li, Na) substitutes for Zn, typically act as deep acceptors and/or are compensated by the donor defects such as oxygen vacancies or interstitial hydrogen \cite{VanDeWalle2000a, Ozgur2005, Anderson2009, McCluskey2015}. Small atoms like Li or Na also self-compensate in the wurtzite structure by acting both as (deep) acceptors when substituting for Zn and shallow donors as interstitial impurities \cite{Park2002, Anderson2009, Lany2010, Lyons2014}.

%
\begin{figure}[t]
\includegraphics[width=\linewidth]{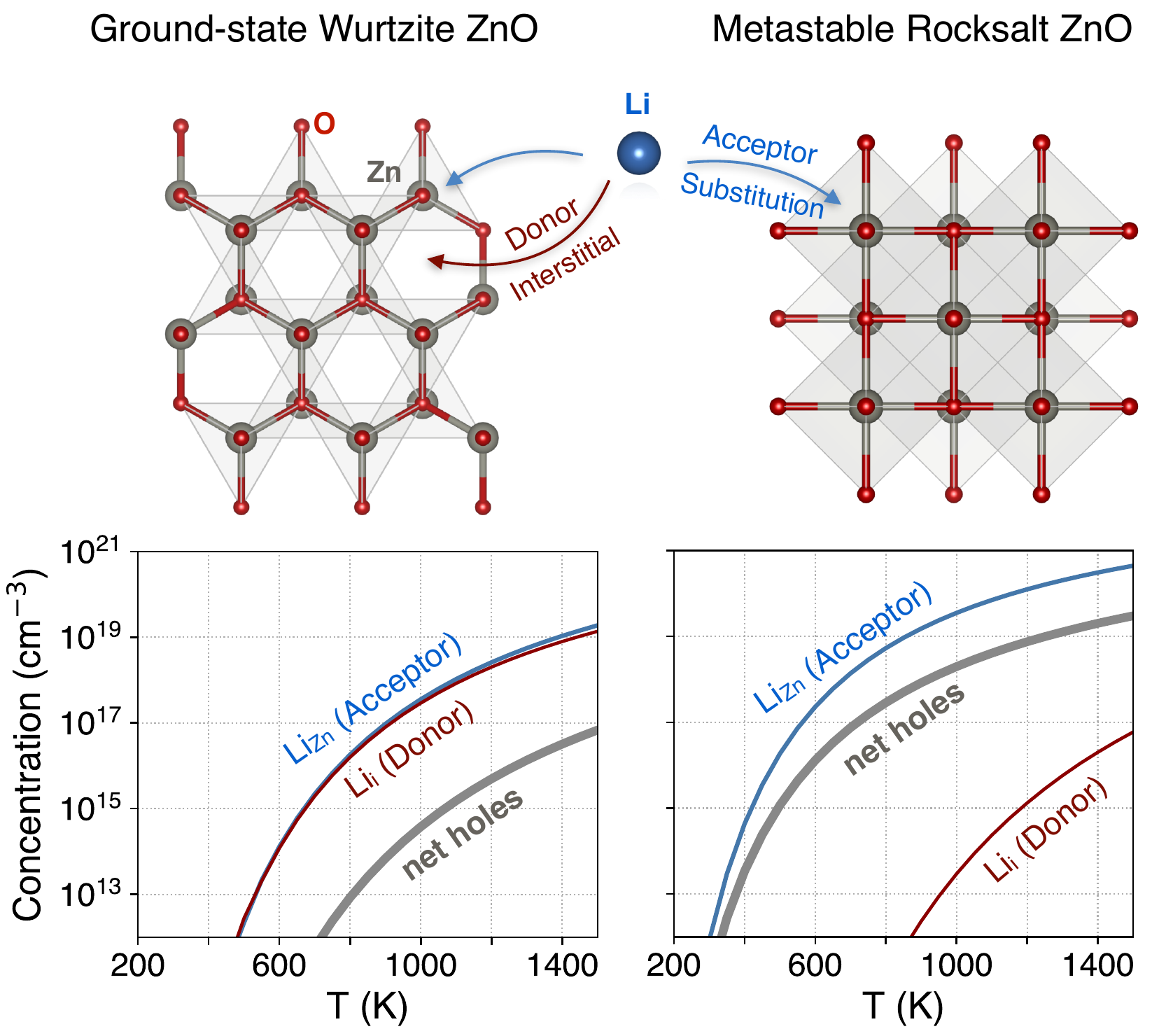}
\caption{\label{fig:1} We predict ZnO to be $p$-type dopable in its metastable, high-pressure rocksalt phase using Li as an external dopant. The ground state wurtzite and the metastable rocksalt structures are shown (upper panel) together with lattice sites Li impurities occupy. In the rocksalt phase Li preferentially substitutes for Zn and acts as a shallow acceptor leading to orders of magnitude higher hole concentrations, shown (lower panel) for both structures at O-rich conditions and as a function of synthesis temperature. 
}
\end{figure}

Here, we investigate whether polymorphism could be helpful in achieving $p$-type doping in ZnO. There are several reasons that make changing the crystal structure a plausible route to overcoming doping bottlenecks of ZnO. First, changing the volume and local coordination could impact the formation of compensating donor defects. In particular, the high-pressure denser phases offer less interstitial space, which could raise the energy to form interstitial defects, such as hydrogen or the group-I dopants. Second and as important, higher energy polymorphs can be expected to have their valence band edge closer to vacuum, which is beneficial for the $p$-type doping \cite{sbzhang_JAP:1998} and for the the formation of shallow acceptor levels rather than deep. This expectation follows from the higher total energy, which includes summation over all occupied electronic states \cite{Ihm1979}.

In addition to the ground state wurtzite, ZnO is known to exists in the metastable zincblende and rocksalt structures \cite{Ozgur2005}. The high-pressure rocksalt phase forms at about $\sim$9 GPa, and can be stabilized at ambient conditions in the nanocrystalline form \cite{Decremps2003, Sokolov2010, Razavi-Khosroshahi2017} or as thin film grown on the MgO substrate with about 15 \% of MgO alloyed \cite{Kunisu2004, Lu2016}. This phase of ZnO is specifically relevant for our discussion because of its octahedral coordination of atoms, smaller volume (about 18  \% reduction) relative to the ground state wurtzite structure and predicted higher valence band edge ($\sim$ 1 eV closer to vacuum) than the wurtzite \cite{Lany2014}. Furthermore, previous demonstration of activated $p$-type transport in rocksalt MgO with Li as an external dopant \cite{Tardio2002}, supports the prospects of the rocksalt structure for alleviating self-compensation of Li impurities. 

\section{\label{sec:II}Results}
We studied the intrinsic defect chemistry, the role of Li as a $p$-type dopant as well as the compensation by unintentional hydrogen in both ground state wurtzite and the metastable rocksalt phase, using modern first-principles defect calculations based on hybrid density functional theory \cite{Freysoldt2014}. Details of the approach are described in the methods section. Confidence in our predictions stems from the correct description of defects and doping in the wurtzite ZnO as well as from our previous works \cite{Goyal2017,Goyal2017c}, in which we demonstrated good agreement between calculated and measured defect and charge carrier concentrations in other systems. As illustrated in Fig.~\ref{fig:1} (left panel), our results agree with the known behavior of Li impurities in the wurtzite structure which simultaneously act as substitutional acceptors and interstitial donors. From the predicted defect concentrations shown in the bottom panel of Fig.~\ref{fig:1}, it is evident that the self-compensation of Li is nearly complete, leading to very low net hole concentrations in the $10^{15}$-$10^{16}$ cm$^{-3}$ range at the growth temperatures well above 1000 K under O-rich/Zn-poor synthesis conditions. These results are consistent with available experimental evidence and previous calculations of Li-doping in wurtzite ZnO \cite{Carvalho2009, McCluskey2015}.

However, in the rocksalt structure, we predict a marked disbalance in the concentration of substitutional and interstitial Li, resulting in the net concentration of free holes to exceed $10^{19}$ cm$^{-3}$ under the same synthesis conditions. The charge carrier (hole) concentrations of this magnitude make the Li-doped rocksalt ZnO a serious contender for the $p$-type transparent conductor applications \cite{Wager2003, Wager2008, Hosono2011a}. As depicted in Fig.~\ref{fig:1} calculated concentration of free holes closely follows that of substitutional Li$_{Zn}$, while the interstitial Li$_{i}$ is predicted to be more than two orders of magnitude less abundant. The difference between the number of Li$_{Zn}$ and the concentration of free holes is due to the presence oxygen vacancies (intrinsic donors) which become more favorable when Fermi energy approaches the valence band and partially compensate the Li$_{Zn}$ acceptors.

%
\begin{figure}[t]
\includegraphics[width=\linewidth]{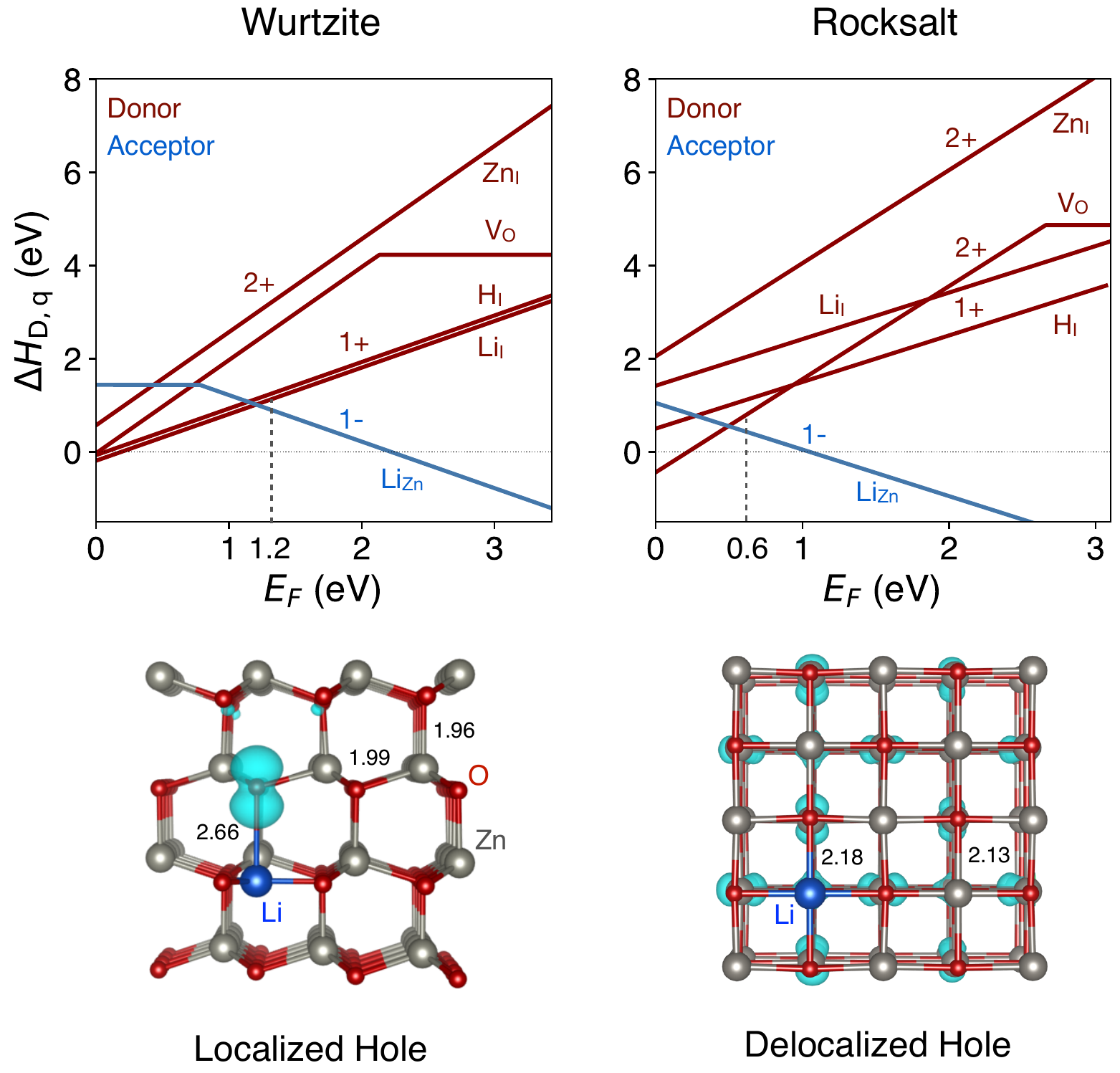}
\caption{\label{fig:2} Calculated defect formation energies at Zn-poor/O-rich conditions as a function of the Fermi energy for both wurtzite and rocksalt ZnO (upper). Only the lowest energy Li$_{\mathrm{Zn}}$ acceptor defect is shown (blue) together with relevant competing donors Li$_{i}$, O-vacancy, Zn$_{i}$, H$_{i}$ (red). Dashed lines mark the position of the equilibrium Fermi energy. (lower) Wavefunctions of the acceptor states are shown for Li$^{0}_{\mathrm{Zn}}$ defect in wurtzite ZnO (localized polaronic state) and rocksalt ZnO (shallow delocalized state).}
\end{figure}

Figure ~\ref{fig:2} (upper panel) presents a comprehensive picture of formation energies of acceptor and dominant donor defects, that forms the basis for predicting defect and charge carrier concentrations. Calculated formation energies of various defects are shown as a function of the Fermi energy. Lines with positive slopes represent donor defects, negative slopes are the acceptors and horizontal lines represent the charge-neutral defects. Energy at which a given defect changes its charge state are typically referred to as the thermodynamic charge transition levels. Dashed lines represent the equilibrium position of the Fermi energy established by the charge neutrality condition between the charged defects and free charge carriers. For clarity we show in Fig.~\ref{fig:2} only the lowest energy acceptor defect Li$_{Zn}$ and the most relevant, potentially compensating, donors including Li$_{i}$, O-vacancies (V$_O$), Zn-interstitials (Zn$_i$) and H-interstitials (H$_i$). A complete picture that comprises all calculated point defects and defect complexes is provided in the supplementary materials along with this paper.

Defect formation energy plots in Fig.~\ref{fig:2} clearly show the compensation between Li$_{Zn}$ and Li$_{i}$ in the wurtzite structure. In addition, the compensating interstitial (unintentional) hydrogen is only 0.26 eV higher in formation energy than Li$_{i}$. Other donors such as V$_O$ and Zn$_i$ are less relevant due to their higher formation energy. Calculated equilibrium Fermi energy is about 1.2 eV above the valence band edge and is only weakly temperature dependent. Hence, the achievable hole concentrations in wurtzite ZnO are very low and are predicted to not exceed $10^{17}$ cm$^{-3}$ even at very high synthesis temperatures. Also, the defect states created by Li$_{Zn}$ are localized, {\it i.e.}, the small polarons with relatively high ionization energy ($\sim0.78$ eV) are predicted to form, which restrict mobility of holes and overall conductivity in already weakly p-type wurtzite ZnO. This behavior is known \cite{Lany2009b, Lany2010, Lyons2014} and well reproduced in our calculations. The localized, polaronic hole state in the wurtzite structure is shown in the lower panel of Fig.~\ref{fig:2}, where the localization of the hole at one of the O atoms neighboring to the Li$_{Zn}$ is evident.

The response of the rocksalt ZnO to Li impurities is markedly different. The formation of Li$_{Zn}$ is clearly more favorable, while both Li$_i$ and H$_i$ are higher in formation energy. Furthermore, the holes created by the substitutional Li are delocalized and formation of small polarons is not favored by the rocksalt structure as shown in Fig.~\ref{fig:2}. We do find localized hole states in the rocksalt structure with localization on the second nearest oxygen, but these are about 0.8 eV higher in energy than the delocalized valence band derived states (see supplementary information). In the rocksalt structure the role of partially compensating donor is taken by O-vacancies, which allow the equilibrium Fermi energy to approach much closer to the valence band. The Fermi energy is about 0.6 eV above the VBM, also weakly dependent on temperature, resulting in free (delocalized) hole concentrations in excess of $10^{19}$ cm$^{-3}$ at synthesis temperatures above 1200 K. Formation of defect complexes such as Li$_{Zn}$+V$_O$, 2Li$_{Zn}$+V$_O$ or Li$_{Zn}$+Li$_{i}$ does not alter the results as shown in the supplementary material.

Another very important feature of the rocksalt ZnO is its bipolar doping nature. Namely, in the absence of Li and at O-poor/Zn-rich conditions rocksalt ZnO remains $n$-type dopable. Our investigation of the intrinsic defects clearly indicates that at O-poor/Zn-rich conditions the absence of low-energy acceptors allows extrinsic $n$-type doping (defects plots provided in the supplementary materials). While the predicted intrinsic electron concentrations of $\sim10^{17}$ cm$^{-3}$ are relatively low, there is no reason to believe that higher electron concentrations could not be attained using external dopants. 

%
\begin{figure}[!t]
\includegraphics[width=0.75\linewidth]{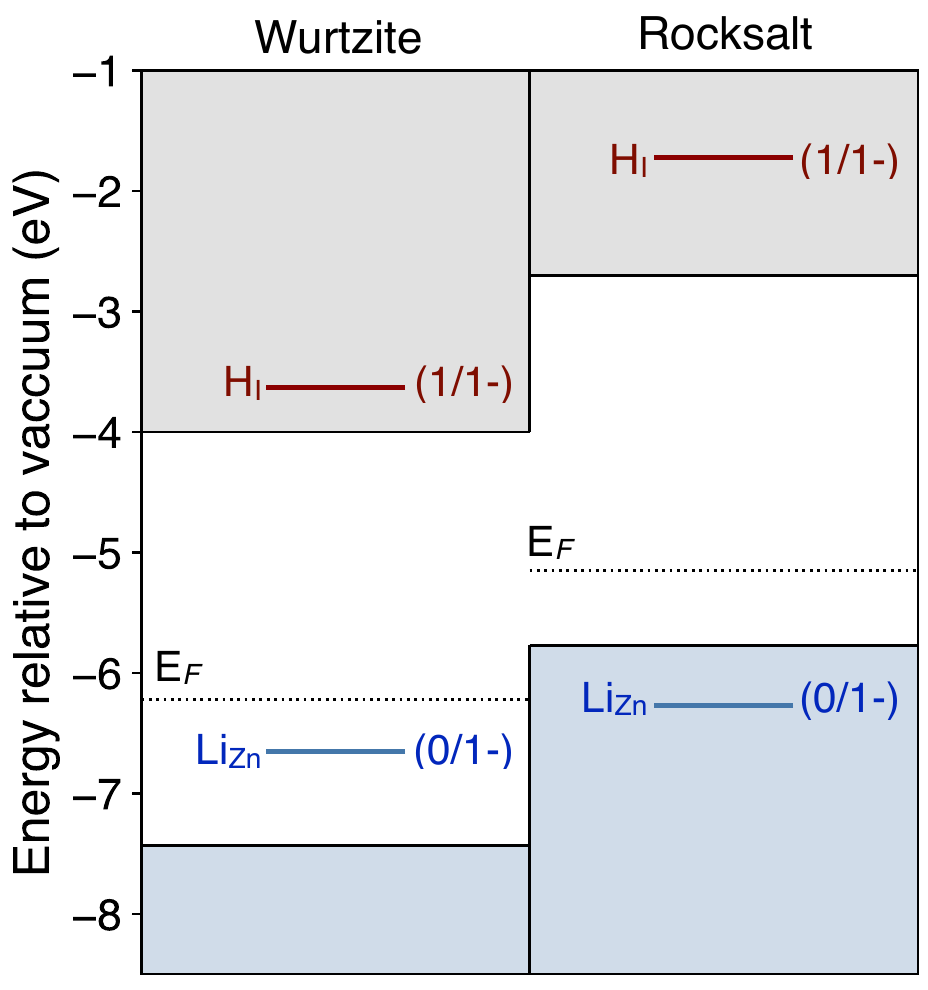}
\caption{\label{fig:3} Comparison of the absolute (relative to vacuum) band-edge positions of the wurtzite (left) and rocksalt (right) phases of ZnO. Charge transition states of substitutional Li and interstitial hydrogen are also shown (thic lines) as well as the the position of the equilibrium Fermi energy (dashed lines) in both structures.}
\end{figure}
%

%
\begin{figure*}[t]
\includegraphics[width=0.875\linewidth]{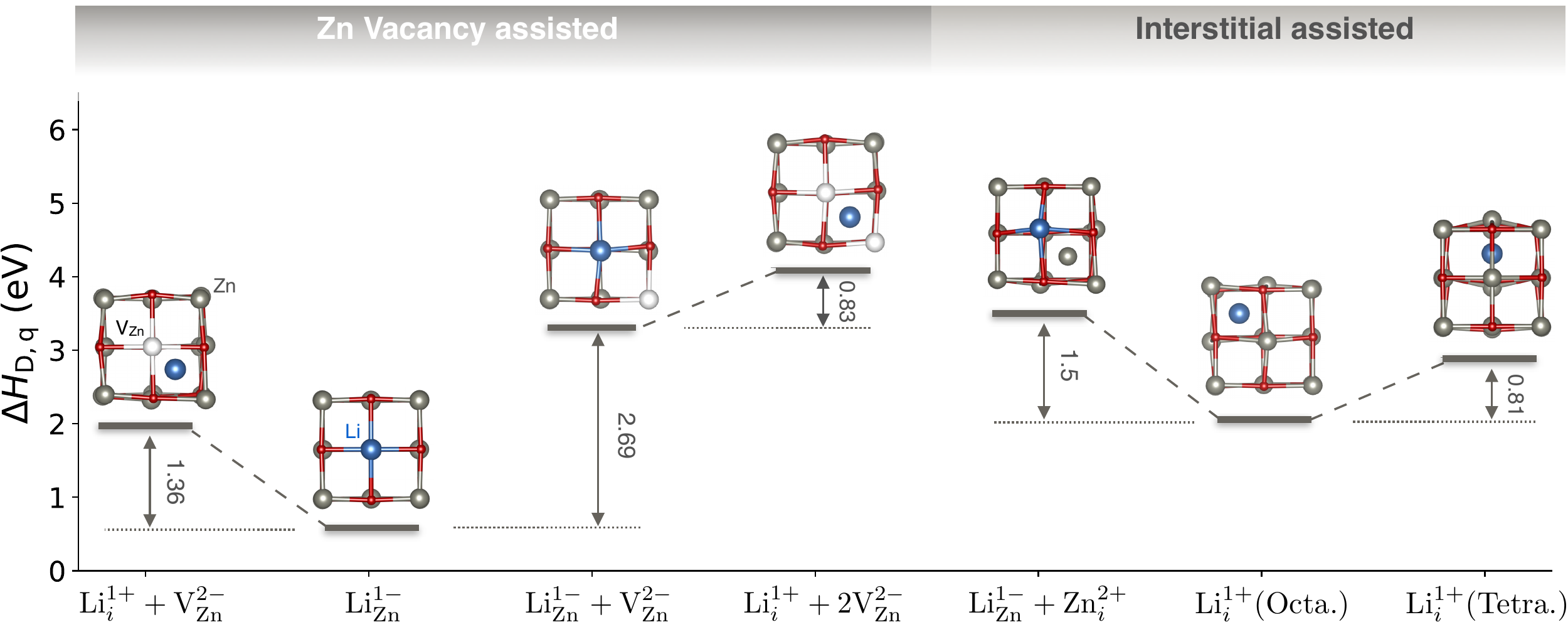}
\caption{\label{fig:4} Formation energies of various defects and defect complexes associated with Li-ion migration in the rocksalt ZnO are shown at the equilibrium Fermi energy under O-rich/Zn-poor growth conditions. Mechanisms are broadly classified into (a) Zn vacancy assisted and (b) interstitial (either Zn or Li) assisted.}
\end{figure*}
%

\section{\label{sec:III}Discussion}

{\it Why is rocksalt ZnO $p$-type dopable?} Per previous discussion, the increase in the formation energy of interstitial compensating donors, such as Li$_i$ and H$_i$, can be attributed in part to the higher density of the rocksalt structure which leaves less interstitial space. However, higher formation energies of {\it charged} interstitial donor defects can also be due to the higher absolute position of the band edges. Indeed, as shown in Fig.~\ref{fig:3} both band edges of the rocksalt phase are closer to vacuum than those of the wurtzite phase, VBM is calculated to be by approximately 1.6 eV higher and CBM is by 1.3 eV.  Nevertheless, the effects of the smaller volume can be seen by examining the interstitial hydrogen. If only the band edge position were to contribute, than the 1+/1- hydrogen charge transition level should remain unchanged on the absolute energy scale between the two structures. Figures~\ref{fig:2} and \ref{fig:3} clearly show that this is not the case, and that the H$_i$ charge transition level is closer to vacuum in the rocksalt structure. This behavior can be explained by the higher density of the rocksalt structure, which due to the size difference affects the formation energy of H$_i$ more in the 1- charge state (H$^-$ ion) than in the 1+ state (bare proton). As a result, the H(1+/1-) charge transition level in the denser phase is closer to vacuum. It is important to note that this behavior does not follow the universal alignment of H(1+/1-) level in semiconductors of Van de Walle and Neugebauer \cite{VandeWalle2003}. A closer look reveals that the universality is demonstrated on a set of tetrahedrally bonded semiconductors, all in their ground state structures. Our H(1+/1-) level for the tetrahedrally bonded wurtzite ZnO agrees well with the universal position of about 4.4 eV below vacuum, but this is not the case for the rocksalt ZnO. Furthermore, relatively good alignment of the Li$_{Zn}$(0/1-) charge transition levels between the two structures is consistent with the expectation that the volume change affects more interstitial than the substitutional defects. Therefore, the misalignment of the H(1+/1-) levels might suggest the existence of different universality classes that correspond to different crystal structures and/or local coordination. 

{\it Prospects for application of $p$-type ZnO.} As already noted the $p$-type dopability of rocksalt ZnO as well as its bipolar nature could be transformative in the context of $p$-type transparent conducting oxides, optoelectronics and transparent electronics in general. In addition to dopability, charge carrier mobility and optical transparency are relevant. Concerning the carrier mobility in the rocksalt ZnO, our results do not show reasons for concern. From the band structure calculations, we find VBM to be at the L-point in the Brillouin zone (see supplementary materials) with the relatively low hole band effective mass $m_h \sim 0.4$ (units of the free electron mass), corresponding to the lighter of the two bands that are degenerate at the VBM. The conduction band is more dispersive with the electron band effective mass  $m_e \lesssim 0.2$. Concerning the ionized impurity scattering in the Li-doped rocksalt ZnO, the shallow nature of Li$_{Zn}$ with the localized polaronic state appearing as a resonance deep in the valence band could be considered less detrimental to transport than the deep defects with states in the band gap \cite{Brandt2015}. In regard to the optical transparency, we calculate the band gap of rocksalt ZnO to be indirect and close to 3.1 eV, in good agreement with many-body GW calculations of S. Lany \cite{Lany2014}. Hence, all direct inter-band transitions lay above the visible part of the solar spectrum. Also, from the band structure we infer weak possibility for intra-band transitions for both holes and electrons, as most of the direct transitions from the band edges into the bands either require energies larger than 3 eV or are between O-2p derived states (holes), and are likely to have low oscillator strengths. Additionally, predicted hole concentrations and effective masses imply the plasma frequency in the far infrared, which removes concerns related to the reflection of light due to plasma excitations. 

{\it Possible synthesis routes and stability of Li-doped rocksalt ZnO.} While the rocksalt phase of ZnO has been stabilized at ambient condition \cite{Decremps2003, Sokolov2010, Razavi-Khosroshahi2017, Kunisu2004}, doping it with Li might be challenging. Based on our results, synthesis route that starts by doping the wurtzite phase and then converts it to the rocksalt is not likely to succeed. This is mainly because of the significant concentration of interstitial Li impurities in the wurtzite phase that are expected to become substitutional in the rocksalt, but depending on the thermodynamics of the process, there will be either Zn or Li precipitates left in the sample. Hence, Li needs to be introduced directly into the rocksalt phase during growth, which in the case of bulk synthesis methods needs to be done at elevated pressure. Vacuum deposition and thin films might be a better route, but the influence of MgO alloying on Li-doping and hole localization requires additional investigation. Ideally, one would prefer to synthesize rocksalt ZnO films with the least possible content of MgO.

Once achieved, Li diffusion in the rocksalt ZnO is of concern in relation to the stability of the $p$-type material. Calculated formation energies of various point defects and defect complexes that are likely transition states along various Li diffusion pathways, indicate that the facile diffusion of Li in the rocksalt ZnO is improbable. In Fig.~\ref{fig:4} we provide relative energies of different intermediate local minima along different Li diffusion pathways, {\it i.e.}, the relative energies of various point defects and defect complexes. Similarly to the wurtzite structure in which Li diffusion has been previously studied \cite{Carvalho2009}, the energy differences between different local minima are relatively high. Diffusion of substitutional Li via interstitial sites (leaving behind the Zn vacancy) requires overcoming the energy difference of 1.36 eV. The exchange of sites between Li$_{Zn}$ and the neighboring V$_{Zn}$ requires overcoming at least 0.83 eV, but also requires the formation of a Li$_{Zn}$+V$_{Zn}$ complex which is high in energy. Direct diffusion of interstitial Li would require a minimum of 0.81 eV, but Li$_i$ are higher in energy and much less abundant than the substitutional Li.  

\section{\label{sec:III}Conclusions}
Using modern defect theory we have shown here that the $p$-type doping in ZnO could be achieved in the metastable, high-pressure rocksalt phase with Li as extrinsic dopant. Our predictions show that in the rocksalt phase extrinsic Li preferentially substitutes for Zn, creates shallow acceptor levels and allows hole concentrations in excess of $10^{19}$ cm$^{-3}$ at oxygen-rich growth conditions. The combination of the close-packed structure and the high absolute band edge energies renders ubiquitous compensating donors such as interstitial Li and/or unintentional H, which are detrimental to $p$-type doping in the ground state wurtzite structure, high in formation energy, and are also responsible for the relatively low hole effective mass $m_h \sim 0.4$. Furthermore, we find that at O-poor conditions, rocksalt ZnO remains $n$-type dopable making this material even more interesting because of its bipolarity. However, we do anticipate challenges in the experimental realization of our predictions mainly related to doping of metastable (high-pressure) phases. But, in our view, the return on investment should be worth an effort given the significance of transparent $p$-type rocksalt ZnO and its bipolar nature to optoelectronics and transparent electronics in general. 

\section{\label{sec:IV}Methodology}

The key quantities that describe the intrinsic and extrinsic doping behavior in semiconductors are the point defect formation energies, and their resulting concentrations. We employ a supercell approach to calculate formation energy $\Delta H_{\mathrm{D}, q}$ of a point defect D in the charge state q using the following equation:
\begin{equation}
\begin{split}\label{eq:1}
\Delta H_{\mathrm{D}, q} (E_{F}, \mu) =  & \,\,[E_{\mathrm{D},q} - E_{\mathrm{H}}] + \sum_{i}n_{i} \mu_{i} + \\
&+ qE_{F} + E_{\mathrm{corr}},
\end{split}
\end{equation}
where, $E_{\mathrm{D},q}$ and $E_{\mathrm{H}}$ are the total energies of the supercells with and without the defect, respectively; $\{\mu_{i}\}$ are the chemical potentials of different atomic species describing exchange of particles with the respective reservoirs; $E_{F}$ is the Fermi energy; and $E_{\mathrm{corr}}$ is the correction term that accounts for the finite-size corrections within the supercell approach \cite{Freysoldt2014}. The chemical potential $\mu_{i} = \mu_{i}^{0} + \Delta \mu_{i}$, is expressed relative to the reference elemental chemical potential $\mu_{i}^{0}$, calculated using the FERE approach \cite{Stevanovic2012} (re-fitted for HSE calculations, see supplementary information), and $\Delta \mu_{i}$ the deviation from the reference elemental phase, bounds of which are determined by the thermodynamic phase stability. The defect formation energy then allows thermodynamic modeling of defect and carrier concentrations, computed here using the approach described in Ref.~\citenum{Biswas2009}.

All calculations are performed using the VASP code \cite{Kresse1996a}, hybrid exchange-correlation functional HSE06 \cite{Heyd2003, Heyd2006} and the exchange mixing of $\alpha$ = 0.3 for rocksalt and $\alpha$ = 0.375 for wurtzite ZnO to match their experimental lattice parameters and their band gaps. The DFT portion of the hybrid HSE functional used the generalized gradient approximation of Perdew Burke Ernzerhof (PBE) \cite{Perdew1996}. Electronic cores are described within the projector augmented wave (PAW) method \cite{Blochl1994} as implemented in the VASP code. A 64 atom supercell and a $\Gamma$-centered $2\times2\times2$ k-point mesh \cite{Monkhorst1976} is used for defect calculations of the rocksalt ZnO. Convergence of the results is validated using a 216 atom supercell. In the wurtzite ZnO defect calculations are performed on a 96 atom supercell with a single $\Gamma$ point. Calculations of defect formation energies, charge transition levels and band edge energies relative to the vacuum level are performed using the standard approach described in Refs.~\cite{Goyal2017c, Stevanovic2014a}. Details of methodology, convergence tests, and additional results with HSE and DFT-PBE are summarized in the supplementary materials.

\begin{acknowledgments}
The authors thank Dr. Stephan Lany from National Renewable Energy Laboratory for fruitful discussions regarding the defect and charge carrier calculations and polaron formation in ZnO. The work was supported as part of the Center for the Next Generation of Materials by Design, an Energy Frontier Research Center (EFRC) funded by U.S. Department of Energy (DOE), Office of Science, Basic Energy Sciences. This research used computational resources sponsored by the DOE Office of Energy Efficiency and Renewable Energy and located at the National Renewable Energy Laboratory (NREL).
\end{acknowledgments}

\bibliography{Manuscript}

\end{document}